\documentclass[aps,prl,twocolumn,showpacs,superscriptaddress,floatfix]{revtex4}
\usepackage[dvips]{graphicx}
\usepackage[english]{babel}
\usepackage{color,enumerate,amsmath,amssymb,amsbsy,amscd,bm}
\usepackage{pstricks,epsf,epsfig}

\newcommand{\ket}[1]{|{#1} \rangle}  % ket
\newcommand{\bra}[1]{\langle {#1}|}  % bra
\newcommand{\ketbra}[2]{|{#1}\rangle\langle{#2}|} % ket-bra
\newcommand{\bs}{{\mathbf s}}   % spin s
\newcommand{\bS}{{\mathbf S}}   % spin S

\begin{document}
\preprint{}
\title[]{ Even-Odd Effects of Heisenberg Chains on Long-range Interaction 
			 and Entanglement }
\author{Sangchul Oh}
\email{sangchul@buffalo.edu}
\affiliation{Department of Physics, University at Buffalo,
State University of New York, Buffalo, New York 14260-1500, USA}
\author{Mark Friesen}
%\email{friesen@cae.wisc.edu}
\affiliation{Department of Physics, University of Wisconsin-Madison,
Madison, Wisconsin 53706, USA}
\author{Xuedong Hu}
%\email{xhu@buffalo.edu}
\affiliation{Department of Physics, University at Buffalo,
State University of New York, Buffalo, New York 14260-1500, USA}
\date{\today}
\begin{abstract}
A strongly coupled Heisenberg chain provides an important channel for quantum 
communication through its many-body ground state. Yet, the nature of the 
effective interactions and the ability to mediate long-range entanglement 
differs significantly for chains of opposite parity. Here, we contrast the 
characters of even and odd-size chains when they are coupled to external 
qubits. Additional parity effects emerge in both cases, depending on the 
positions of the attached qubits. Some striking results include (i) the 
emergence of maximal entanglement and (ii) Ruderman-Kittel-Kasuya-Yosida
(RKKY) interactions for qubits attached to an even chain, and (iii) 
the ability of chains of either parity to mediate qubit entanglement that 
is undiminished by distance.
\end{abstract}
\maketitle
%03.67.Lx Quantum computation architectures and implementations
%03.67.Bg Entanglement production and manipulation
%03.67.Mn Entanglement measures, witnesses, and other characterizations
%03.67.Pp Quantum error correction and other methods for protection against
%         decoherence
%75.10.Jm Quantized spin models, including quantum spin frustration
%75.10.Pq Spin chain models
%75.75.-c Magnetic properties of nanostructures

The addition of even a single particle can have a striking effect on a quantum 
system. For example, spin-$1/2$ particles with antiferromagnetic exchange 
couplings tend to anti-align. The ground state of two coupled spins therefore 
has a compensated magnetic moment, with total spin $S=0$. In contrast, 
a three-spin chain has an uncompensated moment, with total spin $S=1/2$. 
Such even-odd parity effects have been observed recently in a number of 
geometries.  For example, Hirjibehedin {\it et al.}~\cite{Hirjibehedin06} 
assembled antiferromagnetic Heisenberg chains of 1 to 10 manganese atoms 
on a copper nitride surface, revealing parity effects through scanning 
tunneling microscopy, while Micotti {\it et al.}~\cite{Micotti06} measured 
the local magnetic moments of Cr$_7$Cd and Cr$_8$ rings using nuclear magnetic 
resonance methods.

Several attempts have been made to utilize strongly coupled spin chains as 
a medium for quantum communication, showing that long-distance entanglement 
can be generated in several types of spin chains~\cite{Venuti06}, and that 
an odd-size Heisenberg chain can act as a spin bus for coupling remote 
qubits~\cite{Friesen07}. However, parity effects of the spin chain itself 
have not yet been clarified in these quantum information applications. 
In view of how dramatically parity can affect the ground state of a group 
of spins~\cite{Hirjibehedin06}, it is important to investigate how parity 
effects compete with and emerge from the short-range exchange interaction 
in the spin-chain Hamiltonian, how the parity of a spin chain affects 
its capacity to mediate qubit interactions, and how these parity effects 
can be manipulated and taken advantage of.  

In this Letter, we demonstrate dramatic even-odd parity effects of Heisenberg 
chains on induced long-range couplings and entanglement between remote qubits 
weakly attached to the chains. The effective interactions are obtained via 
perturbation calculations.  At first order, an odd-size Heisenberg chain acts 
as a central spin to the qubits. The local magnetic moment of the chain at 
the qubit site determines the sign and strength of the effective coupling 
between them. In contrast, at second order, an even-size Heisenberg chain 
mediates an indirect RKKY interaction between any two external 
qubits~\cite{RKKY}. The sign and strength of the interaction are determined 
by the spin-spin correlations within the chain. We show that these disparate 
coupling mechanisms allow the qubit couplings to be tuned from 
antiferromagnetic to ferromagnetic, and present intriguing opportunities 
for fabricating artificial spin lattices and superlattices.

\begin{figure}[htbp]
\includegraphics[scale=1.0]{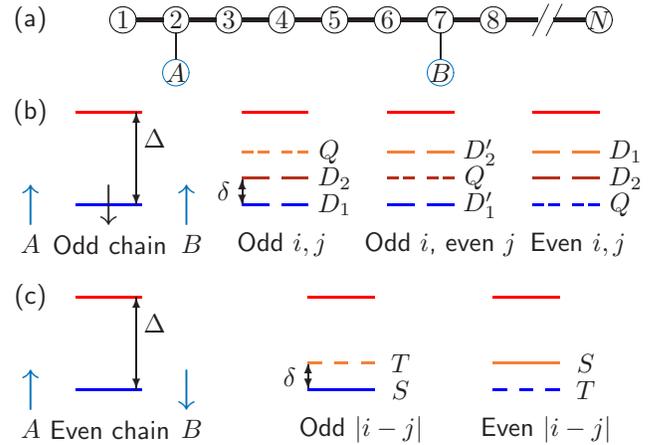}
\caption{(color online). (a) Schematics of two qubits, $A$ and $B$, weakly
attached to nodes $i=2$ and $j=7$ of a chain with $N$ spins.
The qubit perturbation affects the low energy manifolds of (b) an odd-size
chain and (c) an even-size chain.  (Left: unperturbed. Right: perturbed.)
Here $D_1$, $D_2$, and $Q$ denote low-lying doublet and quadruplet states,
respectively. $S$ denotes a singlet and $T$  a triplet.
$\Delta$ is the ground energy gap of the chain. $\delta$ is equal to
$|J^*_{A,i}|=|J^*_{B,j}|$ for an odd-size chain and $|J^*_{A,B}(i,j)|$
for an even-size chain.}
\label{Fig1}
\end{figure}

The model system we consider involves two qubits weakly coupled to 
a Heisenberg chain with $N$ spin-1/2 particles, as shown in 
Fig.~\ref{Fig1}~(a). Such a system can possibly be realized, for example, 
with quantum dot arrays, donor arrays, or magnetic molecules on a surface. 
The total Hamiltonian for the two qubits and the chain is
\begin{align}
H = H_q + H_c  + H_{qc} \,,
\end{align}
where $H_q$ is the Zeeman term of the two qubits and taken to be zero for
simplicity. The Hamiltonian of  the spin chain with antiferromagnetic 
exchange couplings is
\begin{align}
H_c = \sum_{i=1}^{N} J_i\, \bs_i\cdot\bs_{i+1}\,,
\label{Hamil_bus}
\end{align}
where we takes the uniform exchange couplings: $J_i= J_0$ for 
$i=1,\cdots, N-1$.  For open chains, $J_N = 0$,
while for rings, $J_N=J_0$ and $\bs_{N+1} = \bs_{1}$.
A weak antiferromagnetic coupling between qubits, $A$ and $B$, and
chain spins at nodes $i$ and $j$, respectively, is described by
\begin{align}
H_{qc} = J_{A,i}\, {\mathbf S}_A\cdot\bs_{i}
       + J_{B,j}\, \bS_A\cdot\bs_{j}\,,
\label{Hamil_coupling}
\end{align}
where $0< J_{\alpha,i}/J_0 \ll 1$, with $\alpha =A, B$. In this work, 
we consider only two qubits attached to the chain. However, the general 
analysis is applicable to any number of qubits.

We first recall some basic properties of the spin-$1/2$ Heisenberg chain,
defined in Eq.~(\ref{Hamil_bus}).  (i) Partial solutions for $H_c$ can
be obtained via the Bethe ansatz~\cite{Bethe}. (ii) The total spin operator
$\bs = \sum_{i=1}^N\bs_i$ and its $z$ component $s_z=\sum_{i=1}^{N}s_{iz}$
commute with $H_c$, so their eigenvalues are good quantum numbers.
(iii) Finite-size chains exhibit ground state energy gaps that vanish in
the limit $N\rightarrow \infty$.  (iv) The nature of the ground-state 
manifold depends on the even-odd parity as well as the boundary conditions 
for the chain. We now derive the effective interactions when external qubits 
are coupled to such Heisenberg chains.

\paragraph{Odd-size chains.}
An odd-size Heisenberg chain with open boundary has a two-fold degenerate 
ground state, as depicted in Fig.~\ref{Fig1} (b).  The ground doublet
$\{\ket{0_c}, \ket{1_c}\}$ has total spin $s = \hbar/2$ and $s_z = \pm \hbar/2$,
similar to a single spin. The ground states of the odd-size chain can therefore 
be regarded as an extended object, whose spin-$1/2$ character is distributed 
over the entire chain, and is not localized at any given node. 
Thus, when the qubit coupling is weak, the odd-size chain can be effectively 
replaced by a single object called a ``central spin''.

We derive the first-order effective Hamiltonian~\cite{Cohen} by projection:
$H_{\rm eff} = (H_q + H_c)\, P + P\, H_{qc}\, P + {\cal O}( H_{qc}^2)$.
The projection operator $P = \sum_{i,m,n} \ket{i_c,m_A,n_B}\bra{i_c,m_A,n_B}$,
with $i,m,n = 0,1$, spans the qubit eigenstates and the ground state doublet 
of the chain. We thus obtain
\begin{subequations}
\label{Eff_Hamil_odd}
\begin{align}
H_{\rm eff} = J_{A,i}^*\,\bS_A\cdot\bS_{C} + J_{B,j}^*\,\bS_B\cdot\bS_{C}\,,
\end{align}
where the central spin operator $\bS_C$ acting on the ground doublet is 
defined by 
$S_{C,z} = s_z = \frac{\hbar}{2}\left(\ketbra{0_c}{0_c} -\ketbra{1_c}{1_c}\right)$,
$S_{C,x} = \frac{\hbar}{2} \left( \ketbra{0_c}{1_c} + \ketbra{1_c}{0_c} \right)$,
and 
$S_{C,y} = \frac{\hbar}{2} \left( -i\ketbra{0_c}{1_c} + i\ketbra{1_c}{0_c} \right)$.
The effective couplings $J_{\alpha,i}^{*}$ between the central spin and the 
qubits are given by
\begin{align}
\frac{J_{\alpha,i}^{*}}{J_{\alpha,i}} = \bra{0_c}\sigma_{iz}\ket{0_c}
=- \bra{1_c}\sigma_{iz}\ket{1_c}
= \bra{1_c}\sigma_{ix}\ket{0_c}\,,
\label{eq:effective}
\end{align}
\end{subequations}
where Pauli matrices ${\bm\sigma}_i$ act on the $i$-th spin in the chain.

\begin{figure}[ht]
\includegraphics[scale=1.0]{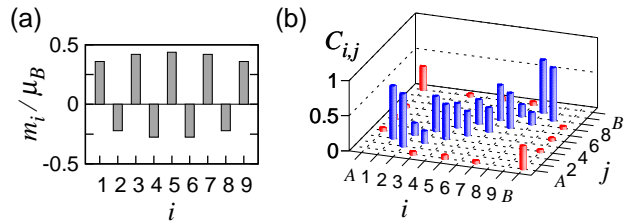}
\caption{(color online). (a) Local magnetic moment
${m_i}/{\mu_B} =\bra{0_c}\sigma_{iz}\ket{0_c}$ of an odd-size chain with $N=9$.
(b) Concurrence $C_{i,j}$ of all possible spin pairs when qubits $A$ and $B$
are coupled to sites $1$ and $9$ of the chain with $N=9$.}
\label{Fig2}
\end{figure}

Equation~(\ref{Eff_Hamil_odd}) shows that qubits $A$ and $B$ may be coupled 
over long distance via the central spin $C$. Furthermore, the effective 
coupling $J_{\alpha,i}^*$ of qubit $\alpha$ to the central spin is given by 
the product of the bare coupling $J_{\alpha,i}$ and the local magnetic moment
$m_i/\mu_B = \bra{0_c}\sigma_{iz}\ket{0_c}$ of the chain at the position where
the qubit is attached. As shown in Fig.~\ref{Fig2} (a), the local magnetic
moments alternate, causing $J_{\alpha,i}^{*}$ to alternate as well. 
Thus, the antiferromagnetic (ferromagnetic) character of the effective 
central-spin-qubit interaction is determined by where the qubit is attached.

There are three possible combinations for the signs of $J^*_{A,i}$ and 
$J^*_{B,j}$, as determined by the parity of the attachment sites $i$ and $j$: 
(i) both positive ($i$ and $j$ both odd), 
(ii) one positive and one negative ($i$ and $j$ with opposite parity), 
and (iii) both negative ($i$ and $j$ both even). Let us rewrite 
Eq.~(\ref{Eff_Hamil_odd}) in the qubit-central spin-qubit (ACB) basis as
\begin{align}
H_{\rm eff}
&= J^{*}_{A,i}\left(\, \bS_A\cdot\bS_{C}
                     + \lambda\,\bS_B\cdot\bS_{C}\,\right)\,,
\label{Eff_Hamil_odd2}
\end{align}
where the magnitude of $\lambda \equiv J_{B,j}^*/J_{A,i}^*$ can be tuned by
the bare couplings $J_{A,i}$ and $J_{B,j}$ under the condition
$J_{\alpha,i} \ll J_0$. We now analyze each of these scenarios.

\begin{figure}[ht]
\includegraphics[scale=1.0]{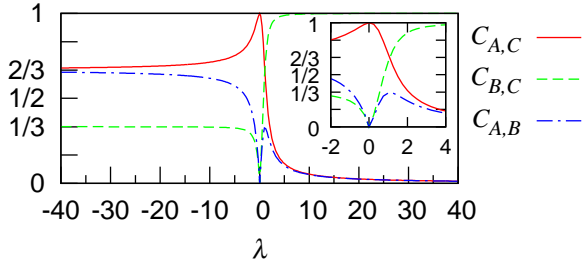}
\caption{(color online). Concurrence between qubits $A$, $B$, and the central 
spin $C$ as a function of $\lambda$, as defined in Eq.~(\ref{Eff_Hamil_odd2}).}
\label{Fig3}
\end{figure}

Case (i), where $\lambda>0$. As depicted in Fig.~\ref{Fig1}~(b), 
the low-energy level ordering is given here by $D_1-D_2-Q$. 
For the special case $J_{A,i}^* = J_{B,j}^* >0$, $\lambda=1$. 
The resulting ground-state doublet can be expressed as
\begin{align}
\ket{G_1}_{ACB}
= \frac{1}{\sqrt{6}} \left( \ket{001} -2\ket{010} + \ket{100}\right) \,,
\label{doublet1}
\end{align}
and its spin-flipped counterpart is $\ket{G_2}_{ACB}$.  More generally,
the eigenstates will depend on $\lambda$. We can quantify the entanglement 
in terms of the concurrence measure \cite{Wooters}, obtaining $C_{A,B}=1/3$ 
for either doublet state. Indeed, the concurrence between qubits $A$ and $B$, 
for any quantum superposition of $\ket{G_1}$ and $\ket{G_2}$, has the same 
value. We can further compute the concurrence in any spin pair, including 
the internal nodes of the chain, as shown in Fig.~\ref{Fig2}~(b). 
As noted in Ref.~\cite{Amico08}, only nearest-neighbor spins in the chain 
exhibit a non-zero concurrence. On the other hand, the qubits and the 
central spin have non-zero concurrence, except in the singular (but trivial)
case of $\lambda =0$, as shown in Fig.~\ref{Fig3}. The concurrence between 
$A$ and $B$ takes its peak value of $1/3$ when $\lambda = 1$, 
and vanishes when $\lambda \gg 1$, due to the competition between the qubits 
to form a singlet state with the central spin.  The central spin is more 
entangled with the qubit having the larger antiferromagnetic coupling.

Case (ii), where $\lambda<0$.  We take $J_{A,i}^{*} > 0$, $J_{B,j}^{*} < 0$.
The low-energy states are now ordered $D_1'-Q-D_2'$,
where two doublets $D_1'$ and $D_2'$ differ from $D_1$ and $D_2$.
For $\lambda =-1$, for example, the ground state doublet $D_1'$ is
$\ket{G_1'}_{ACB}
= {\textstyle{\frac{2+\sqrt{3}}{3+\sqrt{3}}}\, \ket{100} -
              \frac{1}{\sqrt{3}}\,\ket{010} -
              \frac{1}{3+\sqrt{3}}\,\ket{001}}
$
and its spin-flipped counterpart is $\ket{G'_2}_{ACB}$.
Fig.~\ref{Fig3} shows that as $\lambda \to -\infty$, $C_{A,B}$ goes to $2/3$.
The ground states becomes
\begin{align}
\ket{G_1'}_{ACB} 
= \frac{1}{\sqrt{6}} \left( 2\ket{100} -\ket{010} - \ket{001}\right) \,.
\label{doublet3}
\end{align}
It is interesting to note that Eqs.~(\ref{doublet3}) and (\ref{doublet1}) are
identical if $A$ and $C$ are exchanged.

Case (iii), where $J_{A,i}^{*} < 0$ and $J_{B,j}^{*} < 0$, and the low energy
states form a quadruplet.  We find that the entanglement among spins $A$, $B$,
and $C$ depends on the specific state, and is therefore not well defined.

\paragraph{Even-size chains or rings.}

The ground state of an even-size chain or ring is non-degenerate, as shown in
Fig.~{\ref{Fig1}~(c)}. Its quantum numbers are $s=0$ and $s_z =0$,
so that the first-order perturbation due to a weakly coupled qubit vanishes,
and the second-order term is the leading order. In this case, the effective
Hamiltonian for the ground state manifold is given by~\cite{Cohen}
\begin{align}
H_{\rm eff} = {\cal E}_0\ketbra{0_c}{0_c}
            + J^{*}_{A,B}(i,j)\, \bS_A\cdot\bS_B\,,
\end{align}
with the induced effective coupling
\begin{align}
J^*_{A,B}(i,j) 
= 2\sum_{n_c\ne 0} \frac{J_{A,i}\,J_{B,j}}{{\cal E}_0 - {\cal E}_n}
  \bra{0_c}\sigma_{i\mu}\ket{n_c}\bra{n_c}\sigma_{j\mu}\ket{0_c} \,.
\label{RKKY_coupling}
\end{align}
Here $\ket{n_c}$ and ${\cal E}_n$ are the excited eigenstates and eigenvalues
of $H_c$. Note that the induced interaction is also isotropic and
index $\mu$ in Eq.~(\ref{RKKY_coupling}) is any of $x,y$, and $z$,
but not the summation convention. Since the effective coupling involves 
virtually excited states of the chain, Eq.~(\ref{RKKY_coupling}) can be 
expressed, alternatively, 
as $J^{*}_{A,B}(i,j) = 2J_{A,i} J_{B,j}\,\widetilde{G}_{i,j}(0)$, where
$\widetilde{G}_{ij}(\omega)$ is the Fourier transform of the time-dependent 
spin-spin correlation function for the ground state,
$G_{ij}(t) = -i\bra{0_c} \sigma_{i\mu}(t) \sigma_{j\mu} \ket{0_c}$ with
$\sigma_{i\mu}(t) = e^{iH_ct/\hbar}\,\sigma_{i\mu}\,e^{-iH_ct/\hbar}$.
While it is difficult to obtain a closed-form expression for $J^*_{A,B}(i,j)$,
an approximate solution is given by
\begin{align}
J^*_{A,B}(i,j) \approx -2\frac{J_{A,i}\,J_{B,j}}{\Delta}
                     \bra{0_c}\sigma_{i\mu}\sigma_{j\mu} \ket{0_c} \,,
\label{Approx}
\end{align}
where $\Delta= {\cal E}_1- {\cal E}_0$ is the ground energy gap of the chain
and $\sum_{m\ne 0} \bra{0_c} \sigma_{i\mu} \ket{m} \bra{m}
\sigma_{j\mu} \ket{0_c} = \bra{0_c} \sigma_{i\mu} \sigma_{j\mu} \ket{0_c}$
may be used since $\bra{0_c}\sigma_{i\mu}\ket{0_c} = 0$.

We have computed $J^*_{AB}(i,j)$, numerically, by two different methods.
In the first method, we evaluate Eq.~(\ref{RKKY_coupling}) using 
the unperturbed eigenvalues and eigenstates of the chain. 
In the second method, we diagonalize the full Hamiltonian (chain plus qubits), 
and compute the energy gap $\delta$ between the ground state and 
the lowest excited state, all within the manifold of the chain ground state.
For odd qubit separations $|i-j|$ in Fig.~\ref{Fig1}~{(c)},
the ground state of the full system (chain plus qubits) is found to be 
a singlet, while the first excited state is found to be a triplet.
The singlet-triplet energy gap $\delta$ is then given by $J^*_{A,B}(i,j) > 0$,
just as if the two qubits experienced a direct, antiferromagnetic coupling.

\begin{figure}[htbp]
\includegraphics[scale=1.0]{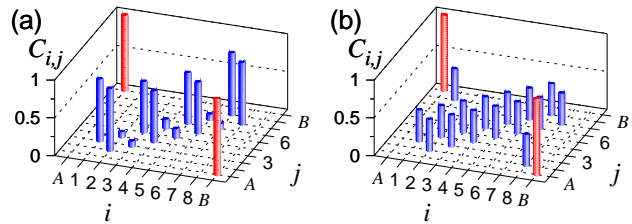}
\caption{(color online). Concurrence $C_{i,j}$ of arbitrary spin pairs
in the ground state of two even-size geometries ($N=8$). (a) Qubits $A$ and
$B$ are attached to sites $1$ and $8$ of an open chain. (b) $A$ and $B$ are
attached to sites 1 and 4 of a ring.}
\label{fig4}
\end{figure}

After tracing out the chain degrees of freedom in the ground state of 
the full system, the reduced density matrix of qubits $A$ and $B$ is almost 
identical to a singlet
$\ket{\psi^{(-)}}_{AB} = \frac{1}{\sqrt{2}}(\ket{01} - \ket{10})$.
The two qubits are therefore maximally entangled, with concurrence 
$C_{A,B} \simeq 1$. The monogamous property of entanglement dictates that
when qubits $A$ and $B$ are maximally entangled, they cannot also be 
entangled with the chain. Therefore, the ground state of the total system is 
approximately the product of the singlet state of the two qubits and 
the ground state of the chain, 
$\ket{\Psi_0} \approx \ket{\psi^{(-)}}_{AB} \otimes \ket{0_c}$.
Furthermore, the concurrence $C_{A,B}$ does not depend on the qubit coupling
locations $i$ and $j$, provided the separation distance is odd. This is 
in contrast to the effective interaction, which decays slowly with qubit
separation, as discussed below. 
In Fig.~\ref{fig4}, we compute the concurrence between spin pairs in two 
different geometries, for the case of odd qubit separation. We observe 
that pair-wise concurrence is translationally invariant for a ring geometry
but not for an open chain.

For an even separation distance $|i-j|$ between the two qubits, the effective
coupling is still determined by the ground state gap, 
$|J_{A,B}^*(i,j)| = \delta$. However, the coupling is now ferromagnetic,
with $ J_{A,B}^*(i,j) < 0$. The ground state of the full system is the product
of a triplet state of qubits $A$ and $B$ and the ground state of the chain,
$\ket{\Psi_0} \approx \ket{T}_{AB} \otimes \ket{0_c}$. The entanglement 
between the qubits depends on a specific superposition of the triplet.

\begin{figure}[htbp]
\includegraphics[scale =1.0]{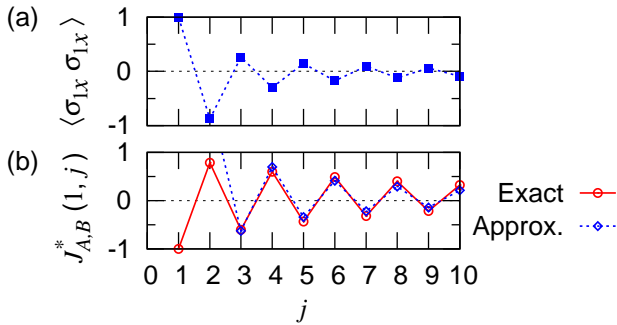}
\caption{(color online). (a) Spin-spin correlation
$\bra{0_c}\sigma_{1x}\,\sigma_{jx}\ket{0_c}$ for the ground state of the
even-size chain with $N=10$. (b) The RKKY interaction $J^{*}_{A,B}(1,j)$
normalized by $|J^{*}_{A,B}(1,1)|$. The labels ``Exact'' and ``Approx.''
correspond to Eqs.~(\ref{RKKY_coupling}) and (\ref{Approx}), respectively.}
\label{RKKY}
\end{figure}

The alternating nature of $J^*_{A,B}(i,j)$ for even and odd qubit separations is
indicative of RKKY effective interactions,
as shown in Fig.~\ref{RKKY}. Note that for an open chain, $J^*_{A,B}(i,j)$
cannot be written as $J^*_{A,B}(|i-j|)$ because of the boundary effect. 
Eq.~(\ref{Approx}) suggests that the RKKY effective coupling $J^*_{A,B}(i,j)$ 
will exhibit a weak power-law decay, similar to the spin-spin 
correlations~\cite{Luther75}.  We can estimate the scaling of the RKKY 
interaction as follows.  The ground state energy gap of the chain is bounded 
by $\Delta \sim \pi^2{J_0}/2{N}$~\cite{Lieb}.
Assuming a weak qubit-chain coupling $J_{\alpha,i} = 10^{-2}\,\Delta$,
we obtain a scaling estimate of $J^*_{A,B}(i,j) \sim 10^{-4}\,\pi^2{J_0}/N$.

\begin{figure}[ht]
%\vspace{8pt}
\includegraphics[scale=0.8,height=2.5cm]{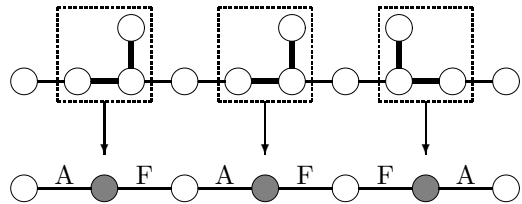}
\caption{Engineered spin superlattices with ferromagnetic and antiferromagnetic
couplings. Strongly coupled three spins (indicated by thick lines) in a dotted
box act as a single effective spin. Couplings between effective spins
and spins can be either ferromagnetic (``F'') or antiferromagnetic (``A''),
depending on attached site.}
\label{Fig6}
\end{figure}

As demonstrated in this Letter, both ferromagnetic and antiferromagnetic 
couplings between qubits can arise in antiferromagnetically coupled chains.
Such tunability provides opportunities to engineer new types of spin structures,
as suggested in Fig.~\ref{Fig6}. As an example, we have sketched a spin 
superlattice with alternating ferromagnetic and antiferromagnetic effective
interactions.  The implications of such novel structures could range from 
new quantum correlations and phases to capabilities in the area of quantum 
information processing. We believe the system considered here can already be 
fabricated and tested with existing nanomagnet or trapped ion 
technologies~\cite{Hirjibehedin06,Micotti06,Zhou10,Kim}, 
while additional tunability could potentially be achieved 
in quantum dot systems.

In conclusion, we have shown that even-odd parity in a Heisenberg chain
produces disparate interactions and entanglement properties between externally
coupled qubits. An odd-size chain acts as a central spin to the qubits,
with effective couplings that are determined by the local magnetic moment of
the chain at the qubit site. On the other hand, an even-size chain mediates
an RKKY interaction directly between the qubits.  While an even-size chain
produces a larger concurrence between the qubits, an odd-size chain provides
stronger couplings and faster gate operations.

\begin{acknowledgments}
This work is supported by the DARPA/MTO QuEST program through a grant
from AFOSR.
\end{acknowledgments}

\end{document}